\documentclass[12pt,leqno]{article}

\usepackage{amsfonts,amssymb,amsthm,amsmath}
\usepackage[mathscr]{eucal}
\usepackage{hyperref}

\advance\textheight\topskip
\newcommand{\mysection}[1]{\section{#1}\setcounter{equation}{0}}

\def\bea{\begin{eqnarray}}
\def\eea{\end{eqnarray}}

\newcommand{\dmn}{d}    

\newcommand{\FQ}{C}
\newcommand{\FR}{D}

\newcommand{\p}[1]{|#1|}
\newcommand{\gh}[1]{\mathrm{gh}(#1)}

\newcommand{\dd}{\partial}
\renewcommand{\d}{\partial}

\renewcommand{\geq}{\,{\geqslant}\,}

\newcommand{\inner}[2]{\langle #1{,}\,#2\rangle}
\newcommand{\binner}[2]{%
  {\langle}\kern-4.15pt{\langle}#1{,}\,#2{\rangle}\kern-4.15pt{\rangle}}
\newcommand{\commut}[2]{[#1{,}\,#2]}

\newcommand{\half}{\mathchoice{%
    \ffrac{1}{2}}{\frac{1}{2}}{\frac{1}{2}}{\frac{1}{2}}}

\newcommand{\ffrac}[2]{\raisebox{.5pt}%
  {\footnotesize$\displaystyle\frac{#1}{#2}$}\kern1pt}

\newcommand{\brst}{\mathsf{\Omega}}

\newcommand{\st}[2]{\overset{#1}{#2}}

\newcommand{\dl}[1]{\mathchoice{\ffrac{\dd}{\dd #1}}{\frac{\dd}{\dd
      #1}}{\ffrac{\dd}{\dd #1}}{\ffrac{\dd}{\dd #1}}}

\newcommand{\bundle}{\boldsymbol}

\def\cE{\mathcal{E}}
\def\cF{\mathcal{F}}
\def\cG{\mathcal{G}}
\def\cH{\mathcal{H}}

\def\cL{\mathcal{L}}

\def\cP{\mathcal{P}}

\def\cS{\mathcal{S}}
\def\cT{\mathcal{T}}

\def\5{\bar }
\def\6{\partial }
\def\7{\hat }
\def\4{\tilde }

\begin{document}
\pagestyle{myheadings}
\markboth{\textsc{\small Barnich, Bonelli, Grigoriev}}{%
  \textsc{\small From BRST to light-cone description of higher spin
    gauge fields}} \addtolength{\headsep}{4pt}


\begin{flushright}
ULB-TH/05-04,\ 
FIAN-TD/05-05\\ 
Sissa 17/2005/EP,\
\texttt{hep-th/0502232}
\end{flushright}

\begin{centering}

  \vspace{0.5cm}

  \textbf{\Large{From BRST to light-cone description of higher spin
    gauge fields}}

\vspace{1.5cm}

\textit{Proceedings of the Workshop ``Quantum Field Theory and
  Hamiltonian Systems'', Caciulata, Romania, 16 - 21 Oct, 2004.}

  \vspace{0.5cm}

  {\large G.~Barnich$^{a}$, G.~Bonelli$^b$ and M.~Grigoriev$^c$}

  \vspace{0.5cm}

\begin{minipage}{.9\textwidth}\small \it \begin{center}
    Physique Th\'eorique et Math\'ematique \\ Universit\'e Libre de
    Bruxelles\\and\\ International Solvay Institutes\\ Campus
    Plaine C.P. 231, B-1050 Bruxelles\\ Belgium \end{center}
\end{minipage}

\end{centering}

\vspace{1cm}

\begin{center}
  \begin{minipage}{.9\textwidth}
    \begin{center} \textbf{Abstract}\\\vspace{10pt}\end{center}
    In this short note we show, at the level of action principles, how
    the light-cone action of higher spin gauge fields can easily be
    obtained from the BRST formulation through the elimination of
    quartets. We analyze how the algebra of cohomology classes is
    affected by such a reduction. By applying the reduction to the
    Poincar\'e generators, we give an alternative way of analyzing the
    physical spectrum of the Fronsdal type actions, with or without
    trace condition.
  \end{minipage}
\end{center}

\vfill

\noindent
\mbox{}
\raisebox{-3\baselineskip}{%
  \parbox{\textwidth}{ \mbox{}\hrulefill\\[-4pt]}} {\scriptsize$^*$
  Senior Research Associate of the National Fund for Scientific
  Research (Belgium). \\[-2pt]
  $^b$ now at: SISSA, High Energy Sector and INFN Trieste, via Beirut
  4, 34014, Trieste, Italy. \\[-2pt] 
  $^c$ now at: Tamm Theory Department, Lebedev Physics Institute,
  Leninsky prospect 53, 119991 Moscow, Russia.}

\thispagestyle{empty} \newpage


\mysection{Introduction}

The title of the lectures given by G.~Barnich during the workshop was
``Higher spin gauge fields: Basics''. The discussion was restricted to
free theories in $4$ dimensions with integer spin. The following
material was covered\footnote{See also the reviews
  \cite{Sorokin:2004ie,Bouatta:2004kk} and the references cited therein.}:

\begin{enumerate}
  
\item Representations of the Poincar\'e group \cite{Cornwell:1985xt},
  \cite{Weinberg:1995mt}

\item Mass-shell field representations \cite{Buchbinder:1998qv}
  
\item Variational principles \cite{Fierz:1939ix}, \cite{Singh:1974qz},
  \cite{Fronsdal:1978rb}
  
\item BRST formulation \cite{Ouvry:1986dv},
      \cite{Bengtsson:1986ys}, \cite{Siegel:1986zi},
  \cite{Henneaux:1987cpbis} 

\item Connection to string theory \cite{Ouvry:1986dv},
  \cite{Bengtsson:1986ys}, \cite{Siegel:1986zi}, \cite{Henneaux:1987cpbis},
  \cite{Bonelli:2003kh}, \cite{Francia:2002pt}, \cite{Sagnotti:2003qa}, 
  \cite{Bekaert:2003uc}
  
\item Trivial pairs and auxiliary fields \cite{Barnich:2003wj},
  \cite{Barnich:2004cr}
 
\item Connection to the light-cone formulation
      \cite{Bengtsson:1983pd},\cite{Siegel:1988yz},\cite{Siegel:1999ew}
  
\item Connection to Vassiliev's formulation 
\cite{Vasiliev:2004qz}, \cite{Barnich:2004cr}

\end{enumerate}

In order to bridge between these lectures and the lectures by L.~Brink
``Light-cone frame formulation of field theories and string theories,
a non-BRST formulation'', we will elaborate on item 7 above and show,
as a pedagogical exercice, how the light-cone action can be explicitly
reached from the BRST formulation. 

More precisely, after reviewing the BRST formulation of the Fronsdal
action, we recall that in BRST language, the reduction to the
light-cone gauge corresponds to the elimination of quartets
\cite{Aisaka:2004ga}. The free light-cone action for higher spin gauge
fields \cite{Bengtsson:1983pd} is then recovered directly from the
reduced BRST operator. We develop in general terms the reduction
of the Lie algebra of BRST cohomology classes under the elimination of
trivial pairs. As an application, we recover the expression of the
Poincar\'e generators \cite{Bengtsson:1983pd} in the light cone gauge.
Finally, we use these expressions in $4$ dimensions to give an
alternative proof of the fact that the Fronsdal type action at level
$s$ describes 

(i) massless particles of helicities $-s,-s+2,\dots,s-2,s$ if no
  trace constraint is required,

(ii) massless particles of helicity $\pm s$ if the trace constraint
  is imposed \cite{Fronsdal:1978rb,Curtright:1979uz,deWit:1979pe}.

\mysection{BRST formulation of the Fronsdal Lagrangian}

We will mostly follow the notations and conventions of \cite{Barnich:2004cr},
to which we refer for further details.

\subsection[Degrees of freedom, constraints and BRST charge]{Degrees
  of freedom, constraints and BRST charge}
\label{charge}

The variables are $x^\mu,p_\mu,a^\mu,a^{\dagger\mu}$, where
$\mu=0,\dots,\dmn-1$.  Classically, $x^\mu$ and $p_\mu$ correspond to
coordinates on $T^*{\mathbb R}^{d-1}$ and $a^\mu,a^{\dagger\mu}$
correspond to internal degrees of freedom.  After quantization, they
satisfy the canonical commutation relations
\begin{equation}
  \commut{p_\nu}{x^\mu}=-\imath\delta^\mu_\nu,\qquad
  \commut{a^\mu}{a^{\nu\dagger}}=\eta^{\mu\nu},
\end{equation}
where $\eta_{\mu\nu}=diag(-1,1,\dots,1)$. 
We assume that $x^\mu,p_\mu$ are Hermitian,
$(x^{\mu})^{\dagger}=x^\mu$, $p_{\mu}^{\dagger}=p_\mu$, while $a^\mu$
and $a^{\dagger\mu}$ are interchanged by Hermitian conjugation.

The constraints of the system are
\begin{equation}
  \cL\equiv\eta^{\mu\nu}p_\mu p_\nu=0,\quad
  \cS\equiv p_\mu a^\mu=0,\quad
  \cS^\dagger\equiv p_\mu a^{\dagger\mu}=0.
\end{equation}
For the ghost pairs $(\theta,\cP)$, $(c^\dagger,b)$, and
$(c,b^\dagger)$ corresponding to each of these constraints, we take
the canonical commutation relations in the form\footnote{We use the
  ``super'' convention that $(ab)^\dagger=(-1)^{\p{a}\p{b}}b^\dagger
  a^\dagger$.}
\begin{equation}
  \commut{\cP}{\theta}=-\imath,\qquad \commut{c}{b^\dagger}=1,\qquad
  \commut{b}{c^\dagger}=-1.
\end{equation}
The ghost-number assignments are
\begin{equation}\label{target-gh-n}
  \gh{\theta}=\gh{c}=\gh{c^\dagger}=1,\qquad
  \gh{\cP}=\gh{b}=\gh{b^\dagger}=-1.
\end{equation}
The hermitian BRST operator is then given by
\begin{equation} \label{eq:Fbrst}
  \brst=\theta\cL+c^\dagger\cS+\cS^\dagger
  c-\imath\cP c^\dagger c,
\end{equation}
and satisfies $\brst^2=0$.

\subsection[Representation space]{Representation space}
\label{sec:space}

Contrary to reference \cite{Barnich:2004cr}, we now choose the
momentum instead of the coordinate representation for the operators
$(x^\mu,p_\nu)$ and also\footnote{This step will allows us to write
  rather compact formulas below, but is not really necessary for the
  analysis.} the holomorphic instead of the Fock representation for
the oscillators $(a_\mu,a^\dagger_\mu), (c,b^\dagger), (b,c^\dagger)$,
while the pair $(\theta,\cP)$ continues to be quantized in the
coordinate representation.

The ``Hilbert space'' $\cH$ consists of the ``wave functions''
$\psi(p,\theta,\alpha^*,\beta^*\gamma^*)= \langle
p\,\theta\alpha^*\beta^*\gamma^*,\psi\rangle$. On these functions, the
operators $p_\mu$, $\theta$, $a^{\mu\dagger}$, $b^\dagger$,
$c^\dagger$ act as multiplication by $p_\mu$, $\theta$,
$\alpha^{\mu*}$, $\beta^*$, $\gamma^*$, while the operators $x^\mu$,
$\cP$, $a^\mu$, $c$, $b$ act as $\imath\frac{\partial}{\partial
  p_\mu}$, $-\imath\frac{\partial}{\partial\theta}$,
$\frac{\partial}{\partial \alpha^{*}_\mu}$, $\frac{\partial}{\partial
  \beta^*}$, $-\frac{\partial}{\partial\gamma^*}$.  The inner product
is defined by
\begin{multline}
  \label{eq:3}
  \langle\psi,\chi\rangle=\int d^dp\, d\theta (\prod_{\mu=0}^{d-1}
  \frac{d\alpha^{\mu*} d\alpha^\mu}{2\pi\imath})d\beta^*d\gamma
  d\gamma^*d\beta\,
\\  \exp{(-\alpha^{\mu*}\alpha_\mu-\beta^*\gamma+\gamma^*\beta)}
{[\psi(p,\theta,\alpha^*,\beta^*\gamma^*)]^*}
\chi(p,\theta,\alpha^*,\beta^*,\gamma^*).
\end{multline}
The string field becomes
\begin{multline}
  \label{eq:2}
  \Psi=\int d^dp\, d\theta (\prod_{\mu=0}^{d-1}
  \frac{d\alpha^*_\mu d\alpha_\mu}{2\pi\imath})d\beta^*d\gamma
  d\gamma^*d\beta\,
\\
  \exp{(-\alpha^{\mu*}\alpha_\mu-\beta^*\gamma+\gamma^*\beta)}
|p\,\theta\alpha\beta\gamma\rangle 
\varphi(p,\theta,\alpha^*,\beta^*,\gamma^*).
\end{multline}

The ghost numbers of the component fields of
$\varphi(p,\theta,\alpha^*,\beta^*,\gamma^*)$ are defined to be 1/2
minus the ghost number \eqref{target-gh-n} of the corresponding state
in the first quantized theory.  The ghost-number-zero component of the
string field can be parameterized as
\begin{equation}
\label{eq:SF-phys}
  \Psi^{(0)}=\Phi-\imath\theta b^\dagger \FQ
  +c^\dagger b^\dagger \FR,
  \end{equation}
  where 
\begin{equation}
\begin{gathered}
  \inner{p\theta\alpha^*\beta^*\gamma^*}{\Phi}=\Phi(p,\alpha^*),\qquad
  \inner{p\theta\alpha^*\beta^*\gamma^*}{\FQ}=\FQ(p,\alpha^*),\\
  \inner{p\theta\alpha^*\beta^*\gamma^*}{\FR}=\FR(p,\alpha^*)\,.
\end{gathered}
\end{equation}

\subsection[Physical and master action]{Physical and master actions}
\label{actions}

The action
\begin{equation}\label{eq:masteraction} 
\bundle{S}[\Psi]=-\half\int
  d^dp\, \inner{\Psi}{\brst \Psi}
\end{equation}
is the solution of the Batalin-Vilkovisky master action associated
with the physical action~
\begin{equation}\label{eq:physaction}
  \bundle{S}^{\mathrm{ph}}[\Psi^{(0)}]=-\half\int  d^dp\,
  \inner{\Psi^{(0)}}{\brst \Psi^{(0)}}.
\end{equation} 

The trace and level
operators
\begin{equation}
\cT=\eta_{\mu\nu}a^\mu a^\nu+2cb\,,
\qquad 
N_s= a^\dagger_\mu
a^\mu-c^\dagger b+b^\dagger c-s \label{eq:7}
\end{equation}
commute with the BRST charge $\brst$ and satisfy $[N_s,\cT]=-2\cT$ so
that they can be consistently imposed as restrictions on the Hilbert
space and on the string field: $\tilde\cH_s={\rm Ker} \cT\cap{\rm Ker}
N_s$, $N_s\tilde\Psi=0=\cT\tilde\Psi$. Assuming appropriate reality
conditions, the restricted action
\begin{equation}\label{eq:physredaction}
  \bundle{S}^{\mathrm{ph}}[\tilde\Psi^{(0)}]=-\half\int  d^dp\,
  \inner{\tilde\Psi^{(0)}}{\brst \tilde\Psi^{(0)}}
\end{equation} 
is the gauge invariant Fronsdal action \cite{Fronsdal:1978rb} 
(written with additional auxiliary
fields contained in $\FQ(p,\alpha^*)$), while
\begin{equation}
  \label{eq:9}
  \bundle{S}[\tilde\Psi]=-\half\int  d^dp\,
  \inner{\tilde\Psi}{\brst \tilde\Psi}
\end{equation}
is the associated proper solution of the Batalin-Vilkovisky master
equation. 

\section[Reduction to the light-cone formulation]{Reduction to the
  light-cone formulation}
\label{sec:reduction-light-cone}

As an alternative to the analysis in \cite{Fronsdal:1978rb},
\cite{Curtright:1979uz} or \cite{deWit:1979pe}, we will analyze the
physical spectrum of actions \eqref{eq:physaction} and
\eqref{eq:physredaction} by going to the light-cone formulation.  In
order to do so, we use the grading introduced in \cite{Aisaka:2004ga}
to relate the BRST to the light-cone formulation in the context of
string theory. This grading allows one to identify quartets for the
full BRST charge on the level of the lowest part in the expansion.

\subsection{Unconstrained reduction}

Here and in the following, indices are lowered and raised with $\eta_{\mu\nu}=
diag (-1,1\dots,1)$. 
The light-cone operators are defined through $p^{\pm}=\frac{1}{\sqrt
  2}(p^0\pm p^{d-1})$, $a^{\pm}=\frac{1}{\sqrt 2}(a^0\pm a^{d-1})$, with
$[a^-,{a^+}^\dagger]=-1=[a^+,{a^-}^\dagger]$ with transverse momenta
and oscillators $p^i,a^i,a^{i\dagger}$, $i=1,\dots,d-2$,
unchanged. In the following, we denote the transverse momenta and
oscillators collectively by a superscript $T$.  The BRST operator
becomes
\multlinegap=1.5cm
\begin{multline}
  \label{eq:4}
  \brst=\theta(p_ip^i-2p^-p^+)+c^\dagger(-p^-a^+-p^+a^-+p_ia^i)+\\
+(-p^-{a^+}^\dagger-p^+{a^-}^\dagger
+p_i{a^i}^\dagger)c-\imath\cP c^\dagger c. 
\end{multline}
\multlinegap=0cm
The grading is defined through the eigenvalues of the operator
\begin{equation}
  \label{eq:4bis}
  G=2{a^-}^\dagger a^+-2{a^+}^\dagger a^--b^\dagger c   - c^\dagger  b, 
\end{equation}
so that 
\begin{equation}
  \label{eq:6}
  \begin{aligned}
    |a^+|&=2, \quad & |{a^+}^\dagger|&=2, \quad & |{a^-}^\dagger|&=-2,
    \quad & |a^-|&=-2,\\ 
    |c|&=1,  \quad & |c^\dagger|&=1,  \quad & |b^\dagger|&=-1, \quad & |b|&=-1,
\end{aligned}
\end{equation}
with all other operators having grading zero. 
The operator $G$ acts on the wave functions
$\phi(p,\theta,\alpha^*,\beta^*,\gamma^*)$ and the associated string
fields as 
\begin{equation}
  \label{eq:5}
  G=-2\alpha^{-*}\frac{\d}{\d\alpha^{-*}}+2\alpha^{+*}\frac{\d}{\d\alpha^{+*}}
-\beta^* \frac{\d}{\d\beta^{*}}+\gamma^*\frac{\d}{\d\gamma^{*}}\,. 
\end{equation}
The Hilbert space $\cH$ can be decomposed according to this grading
and, as the wave functions are polynomial in the
holomorphic variables, each wave function has a component with
lowest degree. The BRST operator decomposes as
$\brst=\sum_{i=-1}^3\brst_i,$ with 
\begin{equation}
  \begin{gathered}
  \label{eq:6bis}
  \brst_{-1}=-p^+(c^\dagger a^-+{a^-}^\dagger c),\qquad
 \brst_0=\theta (-2p^-p^++p_ip^i),\,
\\
 \brst_1=c^\dagger p_ia^i+p_i{a^i}^\dagger c, \qquad
  \brst_2=-\imath \cP c^\dagger c,\qquad  \brst_3=-p^-(c^\dagger
  a^++{a^+}^\dagger c).
\end{gathered}
\end{equation}
If we assume $p^+\neq 0$, which is a standard assumption in the
light-cone approach, states depending non trivially on
$\alpha^{+*},\alpha^{-*},\beta^*,\gamma^*$ form quartets for the
differential $\brst_{-1}$. Indeed, we have 
  \begin{equation}
  \begin{gathered}
    K_1=-\frac{1}{p^+}({a^{+}}^\dagger b-b^\dagger a^+),\qquad
    [K_1,\brst_{-1}]=N_q,\\
N_q= -{a^+}^\dagger{a^-}-{a^-}^\dagger a^+ -c^\dagger b+b^\dagger c,
\end{gathered}
\end{equation}
where the operator $N_q$ acts on states and the associated string
fields as the operator that counts the unphysical variables, 
\begin{equation}
  \label{eq:1}
  N_q=\alpha^{-*}\frac{\d}{\d\alpha^{-*}}+\alpha^{+*}\frac{\d}{\d\alpha^{+*}}
+\beta^* \frac{\d}{\d\beta^{*}}+\gamma^*\frac{\d}{\d\gamma^{*}}. 
\end{equation}

The Hilbert space $\cH$ can accordingly be decomposed as
$\cH=\cE\oplus\cF\oplus \cG$ with $\cG={\rm Im}\,\brst_{-1}$,
$\cF={\rm Im}\,K_1$ and $\cE\subset {\rm Ker}\,\brst_{-1}$
corresponding to the states $\psi(p,\theta,\alpha^{*T})$ that are
eigenvectors of $N_q$ with eigenvalue zero.

Proposition \textbf{3.6} of \cite{Barnich:2004cr} then implies that the full
BRST charge is invertible between $\cF$ and $\cG$: if
$\st{\cF\cG}{\Theta}\equiv(\st{\cG\cF}{\brst})^{-1}$ 
denotes the inverse of the BRST charge
$\st{\cG\cF}{\brst}$ between $\cF$ and $\cG$ and 
$\rho=(\st{\cG\cF}{\brst}_{-1})^{-1}$, we have 
\begin{equation} \label{eq:2bis}
    \smash[b]{
      \st{\cF\cG}{\Theta}=\sum_{n\geq 0}(-1)^n\rho[(\sum_{i\geq 0}
      \st{\cG\cF}{\brst}_i)\rho]^n.}
\end{equation}
Note that the inverse exists because the degree of each wave function is 
bounded from below.  
Furthermore, because the cohomology of $\brst_{-1}$ is concentrated in
degree $0$, the system $(\brst,\cH)$ can be reduced to the system
$(\brst_{0}, \cE)$.  According to proposition \textbf{3.4} of
\cite{Barnich:2004cr}, the associated field theories are related
through the elimination of generalized auxiliary
fields\footnote{Because we are not concerned with issues concerning
  locality of the associated field theories in our discussion, the
  coordinates $x^\mu$ and the momenta $p_\mu$ are considered on the
  same foot with the other degrees of freedom. The discussion of
  \cite{Barnich:2004cr} is then simplified because only the Hilbert
  space $\cH$ is involved in the discussion.}.  In addition, the
elimination of the quartets is compatible with the inner product:
after elimination of the generalized auxiliary fields the inner
  product on $\cH$ reduces to an inner product on $\cE$ that makes
  $\st{\cE\cE}{\brst_0}$ formally self-adjoint. This means that the
  elimination can be done on the level of the corresponding master
  equations, and corresponds to the elimination of generalized
  auxiliary fields in the sense of \cite{Dresse:1990dj}. Explicitly,
  the reduced master action obtained from \eqref{eq:masteraction}
  reads
\begin{multline}
  \label{eq:8}
  \bundle{S}[\Psi^{\cE}]=-\half\int
  d^dp\, \inner{\Psi^{\cE}}{\brst_0 \Psi^{\cE}}
~=
\\
=~-\half \int
  d^dp\,\left(\prod_{i=1}^{d-2} \frac{d\alpha^*_i
    d\alpha_i}{2\pi\imath}\right) \, 
  \exp{(-\alpha^{*i}\alpha_i)}
[\varphi(p,\alpha^{*T})]^*(-2p^-p^++p_ip^i) \varphi(p,\alpha^{*T}).
\end{multline}
As it depends only on ghost number $0$ fields, it coincides with the
physical action and there is no gauge invariance left.

\subsection{Imposing the level and trace constraints}

Because both $\brst_{-1}$ and $K_1$ commute with $N_s$
the elimination can be done consistently also in ${\rm
  Ker}\, N_s$. 

The reduction can also be done in ${\rm Ker}\, \cT$. To see this we
use an expansion with respect to the eigenvalues of $N_q$.  The trace
operator decomposes as $\cT=\cT^{-2}+\cT^0$, with
$\cT^{-2}=-2a^-a^++2cb$, $\cT^0=a^ia_i\equiv \cT^T$, while each wave
function decomposes as $\psi=\psi^0+\dots\psi^M$. Suppose $\cT\psi=0$.
If $\brst_{-1}\psi=0$, we have, for $M\neq 0$, $\psi^M=\brst_{-1}
(\frac{1}{M}K_1 \psi^M)$.  Because $[\cT,K_1]=-\frac{4}{p^+} a^+b$,
the term $K_1\psi$ is double $\cT$ traceless. In fact one can adjust
$K_1\psi$ to make it $\cT$ traceless using the expansion in
homogeneity in all oscilators. Namely, for each monomial $\psi_{(d)}$
of definite homogeneity $d$ in all the oscillators, $\phi_{(d)}\equiv
K_1\psi_{(d)}-k(-2\alpha^{*+}\alpha^{*-}+2\gamma^*\beta^*
+\alpha^{*i}\alpha^*_i)\cT K_1\psi_{(d)}$ is $\cT$ traceless for an
appropriate choice of $k$. Since $\brst_{-1}$ commutes with both $\cT$
and $-2\alpha^{*+}\alpha^{*-}+2\gamma^*\beta^*
+\alpha^{*i}\alpha^*_i$, it follows, for $M\neq 0$, that
$\brst_{-1}(\frac{1}{M}\phi_{(d)})=\psi^M_{(d)}+\dots$, where the dots
denote terms of order strictly lower than $M$. Hence, by summing over
all monomials, $\psi-\brst_{-1}(\frac{1}{M}\phi)$ is $\cT$ traceless,
$\brst_{-1}$ closed and its expansion stops at the latest at $M-1$.
By induction we conclude that $\cT\psi=0=\brst_{-1}\psi$ implies
$\psi=\psi^0+\brst_{-1}\chi$, with $\cT\chi=0$ and $\cT^T\psi^0=\cT
\psi^0=0$.

The same reasoning as in the previous subsection then shows that the
Fronsdal master action \eqref{eq:9} can be reduced to the master and
physical action 
\begin{multline}
  \label{eq:10}
  \bundle{S}[\tilde\Psi^{\cE}]=-\half\int
  d^dp\, \inner{\tilde\Psi^{\cE}}{\brst_0 
\tilde\Psi^{\cE}}~=
\\
=~-\half \int
  d^dp\,\left(\prod_{l=1}^{d-2} \frac{d\alpha^*_l
    d\alpha_l}{2\pi\imath}\right) \, 
  \exp{(-\alpha^{*i}\alpha_i)}
[\tilde\varphi(p,\alpha^{*T})]^*(-2p^-p^++p_ip^i) \tilde
\varphi(p,\alpha^{*T})
\end{multline}
with $\cT^T \tilde\Psi^{\cE}=0=\cT^T\varphi (p,\alpha^{*T})$ and also
$N_s\tilde\Psi^{\cE}=0=N_s\varphi (p,\alpha^{*T})$.

\subsection{Reduction of observables}

By observables, we mean in this context BRST cohomology classes of
operators, i.e., equivalence classes of operators $A$ which commute
with the BRST charge $\brst$, up to such operators which are in the
image of the adjoint action of $\brst$, \bea [\brst,A]=0,\quad A\sim
A+[\brst,B], \eea As discussed for instance in
\cite{Bonelli:2003kh,Barnich:2003wj,Barnich:2004cr}, if one restricts
to antihermitian operators in ghost number zero, these cohomology
classes describe equivalence classes of global symmetries of the
action \eqref{eq:physaction}, with
\begin{eqnarray}
  \delta_{A} \Psi^{(0)}= A \Psi^{(0)}. 
\end{eqnarray}
In order for such observables to be well defined for the constrained
action \eqref{eq:physredaction}, they have to commute with the trace
and the level constraint, 
\bea [N_s,A]=0=[\cT,A].  
\eea

In the following discussion we have in mind in particular the
observables that describe the Poincar\'e transformations. There are
several ways to reduce such observables to the light-cone formulation.
One possibility, discussed in \cite{Bengtsson:1983pd}, is to work out the
compensating gauge transformation needed to stay in the light-cone
gauge when performing the covariant Poincar\'e transformations.
Another possibility, followed in \cite{Henneaux:1987cpbis}, is to use
the quantum Dirac bracket.  Alternatively, one could use a Dirac-type
antibracket for representatives $\langle \Psi,A \Psi\rangle$ of
antifield BRST cohomology classes in ghost number $-1$, which describe
the global symmetry on the level of the master action
\eqref{eq:masteraction}.  This approach has the advantage to extend to
the non linear and non Lagrangian setting and will be discussed
elsewhere. Here we will follow a first quantized BRST approach.

Let $(h_A)\equiv (e_\alpha,f_a,g_a)$ denote a basis for
$\cH=\cE\oplus\cF\oplus\cG$ . Assuming only existence of
$(\st{\cG\cF}{\brst})^{-1}\equiv \st{\cF\cG}{\Theta}$, we can consider
the following invertible change of basis:
\begin{gather}
  \tilde e_\alpha= e_\alpha
  -f_b(\st{\cF\cE}{R})^b_\alpha ,\qquad \st{\cF\cE}{R}=\st{\cF\cG}{\Theta}\,
\st{\cG\cE}{\brst}\\
  \tilde f_a= f_b(\st{\cF\cG}{\Theta})^b_a,\\
  \tilde g_a=
  e_{\beta}(\st{\cE\cG}{L})^\beta_a +
  f_b(\st{\cF\cF}{\brst}(\st{\cF\cG} \Theta)^b_a+ g_a,\qquad
  \st{\cE\cG}{L}=\st{\cE\cF}{\brst}\,\st{\cF\cG}{\Theta}.  
\end{gather}
In other words, $\tilde h=h C$ where 
\bea
C=\left(\begin{array}{ccc} \st{\cE\cE}{\delta} & 0 & \st{\cE\cG} L \\
    -\st{\cE\cE} R & \st{\cF\cG} \Theta &
    \st{\cF\cF}{\brst} \st{\cF\cG} 
\Theta \\ 0 & 0 & \st{\cG\cG} \delta \end{array}\right), \quad 
C^{-1}= \left(\begin{array}{ccc} 
\st{\cE\cE}{\delta} & 0 & -\st{\cE\cG} L \\ \st{\cG\cE}{\brst} &
    \st{\cG\cF}{\brst}  & \st{\cG\cG}{\brst}
    \\ 0 & 0 & \st{\cG\cG} \delta \end{array}\right).
\eea
with $\delta$ denoting the identity matrix. 
The new basis is chosen in such a way that the expression for $\brst$
simplifies to 
\bea
\hat\brst=C^{-1}\brst C=
\left(\begin{array}{ccc} {\tilde\brst} & 0 & 0 \\ 0 & 0 &
    0 \\ 0 & \st{\cG\cF} \delta  &  0 \end{array}\right),
\eea
with 
\bea
{\tilde\brst}=\st{\cE\cE}{\brst}-\st{\cE\cG}{L}\,\st{\cG\cE}{\brst}
-\st{\cE\cF}{\brst}\,\st{\cF\cE}{R}+\st{\cE\cG}{L}\,\st{\cG\cF}{\brst}\,
\st{\cF\cE}{R}=\st{\cE\cE}{\brst}-\st{\cE\cF}{\brst}\,
\st{\cF\cG}{\Theta}\,\st{\cG\cE}{\brst}.
\eea
If $\hat A=C^{-1}A C$, and $A$ is of parity $|A|$,
$[\brst,A]=0$ implies  
\begin{equation}
\hat A=\left(\begin{array}{ccc}{\tilde A} & 0 & 0 \\ 0 & 0 &
    0 \\ 0 & 0  &  0 \end{array}\right)+[\hat \brst,{A_E}],
\label{fund}
\end{equation}
where
  \begin{gather}
    {\tilde A}=\st{\cE\cE}{A}-\st{\cE\cG}{L}\,\st{\cG\cE}{A}
    -\st{\cE\cF}{A}\,\st{\cF\cE}{R}+\st{\cE\cG}{L}\,\st{\cG\cF}{A}\,
    \st{\cF\cE}{R},\label{eq:tildeA}
\\[12pt]
  {A_E}=\left(\begin{array}{ccc} 0 & 0 & (-1)^{|A|}(
        \st{\cE\cF} A -\st{\cE\cG} L \st{\cG\cF} A )\st{\cF\cG} \Theta\\[6pt]
        ( \st{\cG\cE} A -\st{\cG\cF} A \st{\cF\cE}{R}) &
        \st{\cG\cF}{A} \st{\cF\cG} \Theta & (-1)^{|A|} \st{\cG\cH}
        \brst \st{\cH\cF} A \st{\cF\cG} \Theta \\[6pt] 0 & 0 & 0
  \end{array}\right).  
\end{gather}
In particular, 
\bea
 \brst_E=\left(\begin{array}{ccc} 0 & 0 & 0\\
     0 & \st{\cF\cF}{\delta}  & 0 \\ 0 & 0 & 0
  \end{array}\right).
\eea
Equation \eqref{fund} implies that the map $g$ from linear
operators on $\cH$ to linear operators on $\cE$ defined through
$g(A)= \tilde A$ induces a well defined map $g^{\#}$ from the
cohomology of $[\brst,\cdot]$ to the cohomology of
$[\tilde\brst,\cdot]$. It also follows directly from \eqref{fund} that
$g^{\#}([A])=[\tilde A]$ is an isomorphism which preserves the Lie
algebra of observables: $g^{\#}([[A],[B]]=[[\tilde A],[\tilde B]]$.

\subsection{Lorentz generators in the light-cone}\label{sec:Lorentz}

Let $\xi_\mu=\delta_\mu+\omega_{\mu\nu}x^\nu$, with $\omega_{\mu\nu}$
antisymmetric. The Poincar\'e generators are then obtained from the
antihermitian generating operators
\begin{equation}
  \label{eq:11}
  L(\omega,a)=-\imath\big(\xi^\mu p_\mu +
  a^\dagger_\mu\commut{p^\mu}{\xi^\nu} a_\nu\big),
\end{equation}
satisfying
\begin{equation}
  \label{eq:13}
[\brst,L]=0, \quad
[L(\omega_1,\delta_1),L(\omega_2,\delta_2)]=L([\omega_1,\omega_2],
\omega_1\delta_2-\omega_2
\delta_1)
\end{equation}
through
\begin{eqnarray}
  \label{eq:12}
  J^{\mu\nu} & \equiv & -2\imath \frac{\partial L}{\partial
    \omega_{\mu\nu}}=
\big(x^\mu p^\nu-x^\nu p^\mu-\imath ({a^\mu}^\dagger a^\nu
-{a^\nu}^\dagger a^\mu)\big),\\
  P^\mu & \equiv & \imath \frac{\partial L}{\partial \delta_{\mu}}=p^\mu. 
\end{eqnarray}

Now we apply the reduction formula~\eqref{eq:tildeA} to the Poincar\'e
generators in the light-cone basis,
$P^+,P^-,P^i,J^{ij},J^{+i},J^{-i},J^{+-}$. First we note that the
translations and the orbital parts of the Lorentz transformations
carry vanishing degree and map $\cE$ to $\cE$. Moreover all
the terms in~\eqref{eq:tildeA} besides the first one vanish for these
operators because $R$ and $L$ carry strictly positive degree.  

The same reasoning applies to the spin part $S^{ij}=-\imath ({a^i}^\dagger
a^j - {a^j}^\dagger a^i)$ of $J^{ij}$, which are unchanged in the
reduction. The spin operator $S^{+-}=-\imath({a^+}^\dagger a^- -
{a^-}^\dagger a^+)$ also carries vanishing degree, but since its
restriction to $\cE$ vanishes, $\tilde S^{+-}=0$.  The spin operator
$S^{+i}=-\imath({a^{+}}^\dagger a^i-{a^i}^\dagger a^+)$ carries degree $+2$
and therefore $\tilde S^{+i}=0$. Finally, the spin operator
$S^{-i}=-\imath({a^{-}}^\dagger a^i-{a^i}^\dagger a^-)$ has degree $-2$:
therefore the first term in~\eqref{eq:tildeA} vanishes. Furthermore,
the fourth term also vanishes. Indeed, it contains both $R$ and $L$
operators whose expansion contains terms with degree greater or equal
to $2$. This is so because $\Theta$ is of strictly positive degree
while the only possible contributions to $\st{\cG\cE}{\brst}$ and
$\st{\cE\cF}{\brst}$ are coming from $\brst_1$ given
in~\eqref{eq:6bis}. Explicitly, one thus has
\begin{equation}
  \tilde S^{-i} =-\st{\cE\cF}{S^{-i}}\,\rho\,\st{\cG\cE}\brst_1-
\st{\cE\cF}{\brst_1}\,\rho\,\st{\cG\cE}{S^{-i}}\,.
\end{equation}
Choosing $\cG=\brst_{-1} \cH$,
$\cF=\sum_{M>0}\frac{1}{M}K_1\brst_{-1}\cH^M$, where the expansion is
with respect to the eigenvalues of $N_q$, we have $\rho=K_1$
and, for an arbitrary state $\psi^\cE\in\cE$,
\begin{align}
 \rho \st{\cG\cE}\brst_1\psi^\cE&=
\rho c^\dagger p_j a^j\, \psi^\cE=
K_1 c^\dagger p_j a^j\, \psi^\cE=
\frac{1}{p^+}
{a^+}^\dagger p_j a^j\, \psi^\cE \\
\rho \st{\cG\cE}{S^{-i}}\psi^\cE&= 
-\imath \rho {a^-}^\dagger a^i\, \psi^\cE=
-\imath K_1 {a^-}^\dagger a^i\, \psi^\cE=
\frac{\imath}{p^+}b^\dagger a^i\, \psi^\cE\,.
\end{align}
Summing up, one finds
\begin{equation}
  \tilde S^{-i} =\frac{\imath}{p^+}({a^i}^\dagger p^ja_j - 
p^ja_j^\dagger a^i).
\end{equation}

\subsection{Physical spectrum}

We now analyze the particle content of the actions
\eqref{eq:physaction} and \eqref{eq:physredaction} in 4 dimensions.

The Pauli-Lubanski vector is given by
\begin{equation}
  \label{eq:14}
  \tilde W_\mu\equiv \frac{1}{2}\epsilon_{\mu\nu\rho\sigma}
  \tilde J^{\nu\rho}P^\sigma=\frac{1}{2}\epsilon_{\mu\nu\rho\sigma}
\tilde S^{\nu\rho}P^\sigma,\qquad \epsilon_{0123}=-1.
\end{equation}
Using 
\begin{equation}
  \tilde S^{-1}=\frac{\imath p_2}{p^+}({a^1}^\dagger 
a^2-{a^2}^\dagger a^1)\,,\qquad
  \tilde S^{-2}=-\frac{\imath p_1}{p^+}({a^1}^\dagger 
a^2-{a^2}^\dagger a^1)\,,
\end{equation}
and $\epsilon_{+-12}=1$, $\eta_{-+}=-1$ we get
\begin{align}
  \tilde W_+&
=\tilde S^{12}P^- +\tilde S^{-1}P^2-\tilde S^{-2} P^1=
-\imath ({a^1}^\dagger a^2-{a^2}^\dagger a^1)(p_+- \frac{p_\mu p^\mu}{p^+}),\\
\tilde W_1&=\tilde S^{-2}P^+
=-\imath ({a^1}^\dagger a^2-{a^2}^\dagger a^1) p_1\\ 
\tilde W_2&=-\tilde S^{-1}P^+
=-\imath ({a^1}^\dagger a^2-{a^2}^\dagger a^1) p_2\\
\tilde W_-&=-\tilde S^{12}P^+
=-\imath ({a^1}^\dagger a^2-{a^2}^\dagger a^1) p_-
\end{align}

Let us define
\begin{equation}
\alpha^*=\frac{1}{\sqrt 2}(\alpha^{1*}-\imath\alpha^{2*}),\qquad
\bar \alpha^*=\frac{1}{\sqrt 2}(\alpha^{1*}+\imath \alpha^{2*})\,.
\end{equation}
Then the operator $-\imath ({a^1}^\dagger a^2-{a^2}^\dagger a^1)$ acts
on a state $\phi(\alpha^*,\bar\alpha^*)$ as
$\bar\alpha^*\dl{\bar\alpha^*}-\alpha^*\dl{\alpha^*}$.  Reduced fields
$\varphi(p,\alpha^*,\bar\alpha^*)$ of level $s$, $N_s
\varphi(p,\alpha^*,\bar\alpha^*)=0$ can be decomposed as
$\varphi(p,\alpha^*,\bar\alpha^*)= \sum_{k=0}^s(\bar\alpha^*)^k(
\alpha^*)^{s-k} \varphi_{-s+2k}(p)$.  If these fields are on-shell,
$p_\mu p^\mu\varphi_{-s+2k}(p)=0$, the action of the Pauli-Lubanski
vector is given by
\begin{eqnarray}
  \label{eq:17}
  \tilde W_\mu \varphi_{-s+2k}(p)=(-s+2k) P_\mu \varphi_{-s+2k}(p).
\end{eqnarray}
hence, when restricted to level $s$, action \eqref{eq:physaction}
describes massless particles of helicities $-s,-s+2,\dots,s-2,s$. In
addition, because $\cT^T$ acts as $\frac{\partial}{\partial
  \alpha^*}\frac{\partial}{\partial \bar \alpha^*}$, only the fields
$\varphi_{s}(p),\varphi_{-s}(p)$ satisfy the trace condition, so that
action \eqref{eq:physredaction} describes massless particles of
helicities $\pm s$.

\subsection*{Acknowledgments}
The authors want to thank K.~Alkalaev, S.~Cnockaert, A.~Semikhatov and
I.~Tipunin for useful discussions.  The work of Gl.~Ba. and Gi.~Bo. is
supported in part by a ``P{\^o}le d'Attraction Interuniversitaire''
(Belgium), by IISN-Belgium, convention 4.4505.86, by the National Fund
for Scientific Research (FNRS Belgium), by Proyectos FONDECYT 1970151
and 7960001 (Chile) and by the European Commission program
MRTN-CT-2004-005104, in which they are associated to V.U.~Brussel. In
addition, the work of Gi.~Bo.  is partially supported by MIUR.  The
work of M.G. was supported by~RFBR grant 05-01-00996 and by the
grant LSS-1578.2003.2.

\providecommand{\href}[2]{#2}\begingroup\raggedright\endgroup

\end{document}